\documentclass{article}
\usepackage{authblk}

\begin{document}

\title{An attempt to physical science basis of climate changes in early Seventeenth century and the influence of the Little Ice Age in south Italy}

\author[1]{V. Carbone}
\author[2]{L. Parisoli}
\author[2]{R. Cirino}
\author[1]{T. Alberti}
\author[1]{F. Lepreti}
\author[3]{A. Vecchio}

\affil[1]{Dipartimento di Fisica, Universit\'a della Calabria, Ponte P. Bucci Cubo 31C, 87036 Rende (CS), Italy}
\affil[2]{Dipartimento di Studi Umanistici, Universit\'a della Calabria, Ponte P. Bucci Cubo 28B, 87036 Rende (CS), Italy}
\affil[3]{LESIA--Observatoire de Paris, 5 place Jules Janssen, 92190 Meudon, France}

\maketitle

The colder epoch of the Little Ice Age (LIA), defined as Late Antique LIA roughly lasting from 1536 up to 16601, has been characterized by a period of extreme cooling \cite{LIA1,LIA2,LIA3}. Apart for reconstruction of temperature from usual ice cores \cite{icecores}, the effects of LIA are evidenced by looking at localized proxy data, as the late grape harvests in France \cite{grape1,grape2}, the slowing of tree rings growth \cite{LIA1,rings} and progress of glaciers \cite{glaciers}. Since a better geographical coverage with high-resolution and precisely dated reconstructions are presently not available, the synchrony and global significance of LIA has been questioned \cite{ICCP,JIH1,JIH2,JIH3}.

In 1615 Paolo A. Foscarini, a Carmelite monk lived in a monastery of south Italy near Cosenza (Calabria), published a \textit{Trattato} \cite{Trattato} which, at variance to what was common at the time, has not been written in Latin, but in\textit{volgare}, the ancient Italian language. Foscarini is well known as the author of a Epistle in defense of G. Galilei and the Copernican cosmology \cite{Lettera}, addressing the common scriptural objections to the new system of the world. His work was banned in 1616 by the Roman Inquisition, few months before his death. As a consequence, the \textit{Trattato} remained completely forgotten and their content unexploited, until it has been recently rediscovered in the archives \cite{Trattato}. 

We are currently investigating the \textit{Trattato}, and we found strong evidences that it represents, to our knowledge, the first systematic attempt to interpret something unknown at that time, as meteo--climate changes and their forecasting, in the scientific framework of environmental physical effects related to Sun-Atmosphere relationships. 

The intention of Foscarini is clearly evidenced in some sentences of the \textit{Proemio} (Introduction) of the \textit{Trattato} \cite{Trattato}. For example, on page 4 he states:

\textit{Rimane in ogni modo che qui solo si tratti delle Predittioni, \& Antivedimenti delle mutationi de' Tempi, che si causano dalla Causa Materiale, che sono le mere Naturali. [\dots] Noi in questo libro ci prenderemo carrico, che appartiene al Filosofo Naturale, di poter predire e presagire alcuna cosa sopra la Mutatione de' Tempi; dovendo coincidentemente anco trattare, com'egli possa di più prevedere altre cose oltre le Mutationi de' Tempi, naturalmente, per la necessaria connessione, che hanno le Cause naturali con i loro Effetti.}

The sentence can be translated as:

\textit{Here we discuss forecasting and prevision of "weather changes", which are due to Material Causes, which are just Natural Changes. [\dots] In this book, we will undertake something which is in charge of the Natural Philosophe, to predict and foretell something about the weather changes; having to also dealing something else apart for weather changes, due to the natural connection which natural causes have with their effects.}

It is evident that Foscarini, which qualifies themselves as a \textit{Natural Philosophe}, is aimed to investigate weather changes, or, said differently meteo--climatic fluctuations, starting from the hypothesis that they represents natural effects of some environmental changes. In this way, starting from the knowledge of the nature of changes, the author can undertake also additional predictions. These concern social and economical effects of weather and climate changes. This is well stated in another sentence \cite{Trattato}, on page 5, where the author define the utility of the \textit{Trattato}:

\textit{[\dots] si s\'a il quanto importi il prevedere, e prevenire gli accidenti che possono occorrere nella vita humana, per la varia, e sempre instabile vicissitudine, e mutatione de' tempi, e delle stagioni, della quale sogliono sovente venire mille pericoli, e mille disturbi.}

The sentence means:

\textit{[\dots] it is well known how important is to predict and prevent the accidents that may occur in human life due to the constant instability of weather and seasonal changes, from which they often derive a thousand dangers and disorders.}

Social and economic effects, due to extremely cold climate, at that time were superstitiously interpreted, and superstition has started the witch--hunting, because, superstitiously, witches where considered as the cause of all consequences of dangers and disorders caused by climate changes \cite{streghe}. The starting hypothesis of Foscarini, on the contrary, is to avoid superstitious explanation of changes as well as their consequences, as he clearly stated \cite{Trattato} in the question posed on page 18:

\textit{Se delle cose apparenti nel Sole, nella Luna e nelle Stelle, oltre il presagio della mutazione de' tempi, si possano cavare altre naturali predittioni, stando ne i termini della natura, e senza superstitione.}

This sentence represents one of the eight question that Foscarini asks himself as a working hypothesis of the book, namely:

\textit{[I would like to understand \dots] if from appearances in the Sun, the moon and stars, apart for the omens of weather changes, we can get further natural predictions, within natural phenomena, without resorting to superstitious causes.}

The \textit{Trattato} contains, hidden in the Italian language of early seventeenth century and without the aid of a mathematical apparatus, the first historical attempt to naively investigate statistical forecasting of meteo--climatic fluctuations in the framework of complex systems. A detailed systematic analysis of the \textit{Trattato}, in this perspective, will be reported in a forthcoming paper.

To conclude, interestingly, we found some sentences which represent indirect evidences of the influence of LIA in south Mediterranean regions. As an example, on pages 221 of the \textit{Trattato}, Foscarini states \cite{Trattato}:

\textit{Se dop\'o la vindemmia, innanzi il tramontare delle Pleiadi piover\'a, la ricolta senz'altro sar\'a per tempo: Se piover\'a dopo il tramontare delle Pleiadi, sar\'a tarda: Ma se piover\'a insieme con il loro tramontare, sar\'a mediocre, e giusta [\dots] Le Pleiadi [..\dots] a cinque di Novembre tramontano.}

The sentence can be translated as:

\textit{When after the grape harvest, it will rains before the set of Pleiadi, the harvest will be in time: When the rain will be after the set of Pleiadi, the harvest will be late: But in case the rain will happen together with their set, the harvest will be moderate and in time [\dots] The Pleiadi [\dots] set on November, 5.}

To be reliable, the above sentence indicates that, roughly speaking, the date of grape harvest and the set of Pleiadi must be not too far. Since actually in Calabria the grape harvest happens in early September, the sentence represents an evidence of a late grape harvest, on average, in south Italy between the sixteenth and seventeenth centuries. Evidently, at that time, the climate in Calabria must have been colder than the actual climate, thus indirectly confirming the influence of LIA in the south Mediterranean region. Databases of grape harvest dates in France \cite{grape1}, shows that, in the period 1550--1618, the harvests were delayed on average near the middle of October, roughly in agreement with the observation of P.A. Foscarini for south Italy.

\end{document}